\documentclass[12pt]{article}

\def\cos{\c{c}os }

\def\ii{\'{\i }}
\usepackage{graphics}
\usepackage{epsfig}
\usepackage{fancyhdr}
\usepackage{amssymb}
\usepackage{amsmath}
\begin{document}
\title{Elliptic flow at RHIC and LHC in the string percolation approach}

\author{ I. Bautista $^{\dag}$, J. Dias de Deus
\footnote{CENTRA, Departamento de F\ii sica, IST, Av. Rovisco
Pais, 1049-001 Lisboa, Portugal} and C. Pajares \footnote{IGFAE
and Departamento de F\ii sica de Part\ii culas, Univ. of Santiago
de Compostela, 15782, Santiago de Compostela, Spain} }
 \maketitle

\begin{abstract}
The
percolation of strings gives a good description of
the RHIC experimental data on the elliptic flow, $v_{2}$ and predicted a
rise on the integrated $v_{2}$ of the order of $25\%$ at LHC such as it
has been experimentally obtained.
We show that the dependence of $v_{2}$ on $p_{T}$ for RHIC and LHC 
energies is approximately the same as it has been observed, for all the 
centralities. We show the results for different particles and the 
dependence of $v_{2}$ on the centralities and rapidity. Our results are 
compatible with an small value of the ratio $\eta /s$ in the whole 
energy range such as it was expected in the percolation framework.

\end{abstract}

\section{Introduction}
 A major breakthrough was the discovery at RHIC experiments of a
large elliptic flow $v_{2}$ [1-9]. The observed anisotropic flow
can be understood only if the particles measured in the final
state depend not only on the physical conditions realized locally
at their production point but also on the global geometry of the
event. In a relativistic local theory, this non-local information
can only emerge as a collective effect, requiring interactions
between the relevant degrees of freedom, localized at different
points of the collision region. These anisotropic flow is
particularly unambiguous and convincing manifestation of
collective dynamics [10]. The large elliptic flow $v_{2}$ can be
qualitatively explained as follows. In a collision at high energy
the spectator nucleons are fast enough to move away leaving
behind at mid-rapidity and almond shaped azimuthally asymmetric
overlap region filled with the QCD matter. This spatial asymmetry
implies unequal pressure gradients  in the transverse plane, with
a larger density gradient perpendicular to the reaction plane (in-
plane). As a consequence of the subsequent multiple interactions
between the degrees of freedom involved, the spatial asymmetry
leads to anisotropy in the momentum space [4,11,12]. The final
particle transverse momentum is more likely to be in-plane that in
the out-plane, with $v_{2}>0$ as predicted [5].

The general idea has been realized in various mechanisms for the
source of elliptic flow. A convenient and successful way to
describe the flow anisotropy is achieved in the hydrodynamical
approach [6,8], taking into account the asymmetric shape of the
nuclei overlap in the transverse plane at values of the impact
parameter $b$ different from zero. The description of course
assumes collective effects to be responsible of the flow. The
microscopic framework of the hydrodynamics approach, as well as
the understanding of the required early thermalization remain open
questions. The percolation of strings [13,14] can provide the
microscopic picture and the answer of these questions.

In a heavy ion collision at high energy, strings stretched between
the partons of the projectile and the target are formed. Each
string in the transverse space looks like disk of area $\pi
r_{0}^{2}$, with $r_{0}\simeq 0.2-0.3$ fm due to the confinement.
As the energy and/or the size of colliding objects grows the
number of strings increases and the strings overlap forming
clusters, whose behavior is determined by the color field inside
which is given by the vectorial sum of the color field of the
individual string. In this way, the formation of clusters can be
seen as interactions of the partons of the individual strings with
the corresponding color rearrangement. On the other hand, a local
temperature can be defined associated to the fragmentation of the
string via the Schwinger mechanism [15,16], and in the same way a
temperature can be associated to the cluster formed from many
individual strings. As far as this cluster covers most of the
total interaction area, (what happens if the string density is
over the critical percolation threshold) this temperature becomes a
global temperature determined by the string density. This fact, together 
with the good description of the
elliptic flow of RHIC data makes the percolation of strings a good
candidate to be the initial state for hydrodynamical description.

The elliptic flow was studied in the framework of percolation in
references[17-20] leading to the developments of analytical formulaes for 
$v_{2}$
including the dependence on transverse momentum, which were
successfully compared with the RHIC experimental data [17-18]. In
reference [20] were given arguments in favor of the methods used
previously.

In this paper we present our results for LHC energies. Previously,
we already anticipated a rise of around $25\%$ [19] for the
integrated elliptic flow as it has been obtained experimentally
[21]. In the evaluation , the main source of uncertainties is the
string density due to the uncertainties in the number of strings
and its dependence on the energy and centrality. To avoid these 
uncertainties we determine the
string density in $Pb-Pb$ at LHC at different centralities
comparing the corresponding formula in percolation with the data
of the ALICE collaboration [22]. Given the string density, the
elliptic flow is obtained using our analytical formula.

We find that $v_{2}$ for $p_{T}<0.8$ GeV/c is very similar at LHC and
RHIC energies, being slightly larger for $p_{T}>0.8$ GeV/c.
Concerning the rapidity distribution there is no longer a triangle
shape and a flatten distribution is obtained for $|y|<4$. Also
there is not limiting fragmentation in the same way that in
percolation there is not limiting fragmentation in
$\frac{dN}{dy}$. We obtain a very reasonable overall agreement
with the experimental data, compatible with the existence of a
partonic fluid with a low ratio between the shear viscosity and the 
entropy density in the whole energy range from RHIC to LHC. This low
ratio for this energy range was expected in the percolation framework 
[23-24].

The plan of the paper is as follows. In the next section we present a
brief introduction to percolation of strings. In section 3 we
describe the obtention of the analytical formulas for the elliptic
flow. In section 4 we discuss our results comparing with LHC and
RHIC experimental data. Finally, in section 5 we present our
conclusions.

\section{The string percolation model}
In the collision of two nuclei at high energy, the multiparticle
production is described in terms of color string stretched between
the partons of the projectile and the target. These strings decay
into new ones by $q\bar{q}$ or $qq-\bar{q}\bar{q}$ pair production
and subsequently hadronize to produce hadrons. Due to the
confinement, the color of these strings is confined to a small
area in transverse space $S_{1}=\pi r_{0}^{2}$ with $r_{0}\simeq
0.2-0.3$fm. This value corresponds to the correlation length of
the QCD vacuum [25]. With increasing energy and/or atomic number
of the colliding particles, the number of strings grows and they
start to overlap forming clusters. At a certain critical density, a
macroscopical cluster appears, which marks the percolation phase
transition. This value corresponds to the value
$\rho_{c}=1.18-1.5$ (depending on the type of the profile
functions of the nucleus employed, homogeneous or Wood-Saxon)
where the variable $\rho=N_{s} S_{1}/S_{A}$, and $S_{A}$
\begin{equation}
S_{A}=2 R_{A}^{2}[\cos^{-1}(\beta)-\beta \sqrt{1-\beta^{2}}]
\end{equation}
with
\begin{equation}
\beta=\frac{b}{2R_{A}}
\end{equation}
is the overlapping area of the nucleus. In order to describe the
behavior of the clusters formed by several overlapping strings we
must introduce some dynamics. We assume that a cluster of $n$
strings behaves as a single string with an energy-momentum that
corresponds to the sum of the energy-momentum of the individual
strings with a higher color field, corresponding to the vectorial
sum of the color field of each individual string. In this way
[14][26] we can compute the mean multiplicity $\mu_{n}$ and the
mean transverse momentum squared $<p_{T}^{2}>_{n}$ of the
particles produced by a cluster:
\begin{equation}
\mu_{n}= \sqrt{\frac{nS_{n}}{S_{1}}}\mu_{1}, \mbox{ and }
<p_{T}^{2}>_{n}= \sqrt{\frac{nS_{1}}{S_{n}}}<p_{T1}^{2}>
\end{equation}
where $\mu_{1}$ and $<p_{T}^{2}>_{1}$ are the mean multiplicity
and mean $p_{T}^{2}$ of particles produced by a single string and
$S_{n}$ is the total area occupied  by the $n$ disks, which can be
different for a different configurations even if the clusters have
the same number of strings. Note that if the strings just touch
each other, $S_{n}=nS_{1}$, and the strings act independently of
each other. In contrast, if they fully overlap $S_{n}=S_{1}$, and
it is obtained the largest suppression of the multiplicity and the
largest increase of the transverse momentum. In the limit of high
density, one obtains [14][26]
\begin{equation}
<n\frac{S_{1}}{S_{n}}>=\frac{\rho}{1-e^{-\rho}}\equiv
\frac{1}{F(\rho)^{2}}
\end{equation}
and equations (3) transforms into analytical ones
\begin{equation}
\mu=N_{s}F(\rho)\mu_{1} \mbox{,   }
<p_{T}^{2}>_{1}=\frac{<p_{T}^{2}>_{1}}{F(\rho)}
\end{equation}
In the mid-rapidity region, the number of strings is proportional
to the number of collisions $N_{coll}\sim N_{A}^{4/3}$ and grows
like $s^{2\Delta}$. Therefore $\rho \sim N_{A}^{2/3}$, and $\mu\sim
N_{A}$ and $\mu \sim s^{\Delta}$. In other words, the multiplicity
per participant is almost independent of $N_{A}$ (the only
dependence on $N_{A}$ arises from the factor $\sqrt{1-e^{-\rho}}$
in eq. (5)). Outside the midrapidity, $N_{s}$ is proportional to
$N_{A}$ instead of $N_{A}^{4/3}$. Therefore, there is an
additional suppression factor $N_{A}^{1/3}$ compared to central
rapidity.

Concerning the transverse momentum distribution, one needs the
distribution $g(x,p_{T})$ for each cluster and the mean squared
transverse momentum of the clusters, $W(x)$. For $g(x,p_{T})$ we
assume the Schwinger formula $g(x,p_{T}^{2})=exp(-p_{T}^{2}x)$.
For the weight function, we assume the gamma distribution
\begin{equation}
W(x)=\frac{\gamma(\gamma x)^{k-1}}{\gamma \Gamma (k)}e^{-\gamma x}
\end{equation}
which is the simplest distribution among the distributions stable
under the size transformations [27,28]. In (6) $\gamma$ and $k$
are
\begin{equation}
\frac{1}{k}=\frac{<x^{2}>-<x>^{2}}{<x>^{2}} \mbox{,  } \gamma
=\frac{k}{<x>}
\end{equation}
$\frac{1}{k}$ is proportional to the width of the distribution,
depending on the string density $\rho$. At small density, there is
not overlapping of strings and all the strings produce particles
with the same mean transverse momentum. The width is zero and $k
\rightarrow \infty$. When $\rho$ increases there is some
overlapping of strings, different clusters are formed and k
decreases. The minimum of $k$ will be reached when there are more
different clusters. Above this point, the different clusters start
to join and $k$ increases. In the limit $\rho \rightarrow \infty$,
there is only one cluster formed by all the produced strings and
$k \rightarrow \infty $. Therefore the transverse momentum
distribution $f(p_{T},y)$ of the particle i is given by [28]
\begin{equation}
\begin{split}
f(p_{t},y)=\frac{dN}{dp_{T}^{2}dy}=\int_{0}^{\infty}d(x)W(x)g(p_{T},x)\\
=\frac{dN}{dy}\frac{k-1}{k}\frac{1}{<p_{T}^{2}>_{i}}F(\rho)(1+F(\rho)p_{T}^{2}/k<p_{T}^{2}>_{i})^{-k(\rho)}
\end{split}
\end{equation}
where the dependence of $k$ or $\rho$ is determined by comparing
(8) with the experimental data. The formula (8) is valid for all
the energy, centralities and $p_{T}<4-5$ GeV/c. In the case of
baryons we must introduce minor changes [29]. The equation (8) is
the main ingredient for the evaluation of the elliptic flow.

\section{Elliptic flow}

The cluster formed by the strings has generally an asymmetric form
in the transverse plane and acquires dimensions comparable to the
nuclear overlap. This azimuthal asymmetry is at the origin of the
elliptic flow in percolation. In fact, the partons emitted at some
point inside the cluster have to pass a certain length through the
strong color field of the cluster before they appear outside. It
is natural to assume that the energy loss by the parton is
proportional to the length and therefore the transverse momentum
$p_{T}$ of an observed particle will depend on the path length
travelled and so different for different direction of emission
[19].

In this way the inclusive cross section
$\frac{dN}{dp_{T}^{2}dyd\varphi}=f(p_{T}^{2},\rho_{\varphi},y)$
can be obtained from eq (8). doing the change $\rho\rightarrow
\rho_{\varphi}$, where
\begin{equation}
\rho_{\varphi}=\rho(\frac{R}{R_{\varphi}})^{2}
\end{equation}
where $R$ and $R_{\varphi}$ are shown in fig 1. Notice that
\begin{equation}
\frac{\pi R^{2}}{4}=\frac{1}{2}\int_{0}^{\pi/2}d\varphi R_{\varphi}^{2}
\end{equation}
\begin{figure}
\begin{center}
      \resizebox{100mm}{!}{\includegraphics{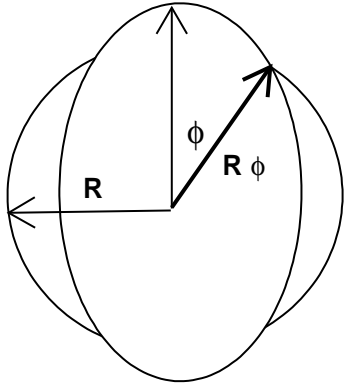}}
    \caption{Azimuthal dependence of R.}
    \label{test4}
\end{center}
\end{figure}

Expanding $f(p_{T},\rho_{\varphi},y)$ around $\rho$ or $R$ we have
\begin{equation}
f(p_{T},\rho_{\varphi},y)\simeq
\frac{2}{\pi}f(p_{T},\rho,y)[1+\frac{\partial\ln
f(p_{T}^{2},\rho,y)}{\partial R^{2}}(R_{\varphi}^{2}-R^{2})]
\end{equation}
Notice that due to (10) we have
\begin{equation}
\int_{0}^{\pi/2}f(p_{T},\rho_{\varphi},y)d\varphi =f(p_{T},\rho,y)
\end{equation}
and from (11), finally we obtain
\begin{equation}
\begin{split}
v_{2}(p_{T}^{2},y)=\frac{2}{\pi}\int_{0}^{\pi /2}d\varphi
cos(2\varphi)[1+\frac{\partial \ln f(p_{T}^{2},\rho,y)}{\partial
R^{2}}(R_{\varphi}^{2}-R^{2})]\\
=\frac{2}{\pi}\int_{0}^{\pi/2}d\varphi \cos 2\varphi
(\frac{R_{\varphi}}{R})^{2}
(\frac{e^{-\rho}-F(\rho)^{2}}{2F(\rho)^{2}})\frac{F(\rho)p_{T}^{2}/<p_{T}^{2}>_{1}}{(1+F(\rho)p_{T}^{2}/<p_{T}^{2}>_{1})}
\end{split}
\end{equation}
having present that
\begin{equation}
\frac{R_{\varphi}}{R_{A}}=\frac{\sin(\varphi-\alpha)}{sin \varphi}
\end{equation}
\begin{equation}
\alpha= \sin^{-1}(\beta \sin \varphi)
\end{equation}
for fixed $\sqrt{s}$ and $\rho$, in the limit cases of $b
\rightarrow 0$ and $b \rightarrow$ $2R_{A}$ we have:

$(i)$ $b \rightarrow 0$ (i.e $\beta \rightarrow 0$ or $N_{A}
\rightarrow A$) which implies $\alpha \rightarrow 0$, and
$v_{2}(p_{T}^{2},y)\rightarrow 0$.

 $(ii)$ $b \rightarrow 2R_{A}$
(i.e $\beta \rightarrow 1$ or $N_{A} \rightarrow 0$) which implies
$\alpha  = \varphi$ and $v_{2}(p_{T}^{2},y) \rightarrow 0$ if we
look now to $p_{T}^{2}$ dependence of $v_{2}$, we see that $v_{2}
\rightarrow 0$ as $p_{T}^{2}\rightarrow 0$ and $v_{2} \rightarrow$
constant as $p_{T}^{2} \rightarrow \infty$.

 We observe that at low
$p_{T}$, say $p_{T}<1$, the dependence on $\rho$ of (13) is given
by $(e^{-\rho}-F(\rho)^{2})/2F(\rho)$ which remain approximately
constant between $\rho=2$ and 5 which corresponds to central
$Pb-Pb$ collisions at RHIC and LHC energies, therefore we expect
the same $v_{2}$ at low $p_{T}$ at RHIC and LHC energies.

We perform next the integration in $p_{T}^{2}$, weighted by
$\frac{dN}{dp_{T}^{2}dy} / \frac{dN}{dy}$ to obtain
\begin{equation}
v_{2}=\frac{2}{\pi}\int_{0}^{\pi/2}d\varphi cos(2\varphi)
(\frac{R_{\varphi}}{R})(\frac{e^{-\rho}-F(\rho)^{2}}{2F(\rho)^{3}})\frac{R}{R-1}
\end{equation}
Again, we have $v_{2}\rightarrow 0$ as $N_{A} \rightarrow A$ and
also $v_{2}\rightarrow 0$ as $N_{A}\rightarrow 0$. As the bracket
in (16) is in modulus a growing function of $\rho$, $v_{2}$ at
fixed rapidity is a growing function of energy and $N_{A}$

In order to obtain formula (13) we have retained only the first term of 
the expansion on $R^{2}$. This is a reasonable approximation at low 
$p_{T}$. For higher $p_{T}$ we expect corrections of the order of 
$10-15\%$.

\section{Results and discussion}
 The first formula in 
(5) can be written as
\begin{equation}
\sqrt{(1-exp(-\rho))\rho}=\frac{\pi
r_{0}^{2}}{\mu_{1}}\frac{1}{S_{A}}\frac{dn}{d\eta}
\end{equation}
As we know $r_{0}=.2$ fm and $\mu_{1}=.8$ obtained previously by
comparing eq. (17) with $pp$ data, we can compute the values of
$\rho$ from eq. (17). This allows us to eliminate the main source
of uncertainties which comes from the value of the number of
strings $N_{s}$ and its dependence on the energy and on the
centrality. We know that $N_{s}\sim s^{2 \Delta}$ and $N_{s}\sim
N_{A}^{4/3}$ and we obtained $2\Delta\simeq 2/7$ from
energy-momentum conservation arguments [34], but in this paper, we
determine the string density directly from the experimental data
on $\frac{dN}{d\eta}$ to eliminate any uncertainty. From the
centrality dependence of the experimental data [22] at
$\sqrt{s}=2.76$ TeV for $Pb-Pb$ collisions and at $\sqrt{s}=200$
GeV for $Au-Au$ collisions we obtained the dependence of $\rho$
on the centralities at these two energies as it is shown in Fig. 
(2). In 
Fig. (3) we
show the obtained centrality dependence of
$\frac{2}{N_{A}}\frac{dN}{d\eta}$ with the $\rho$-values together
with the experimental data [22].
\begin{figure}
\begin{center}
      \resizebox{100mm}{!}{\includegraphics{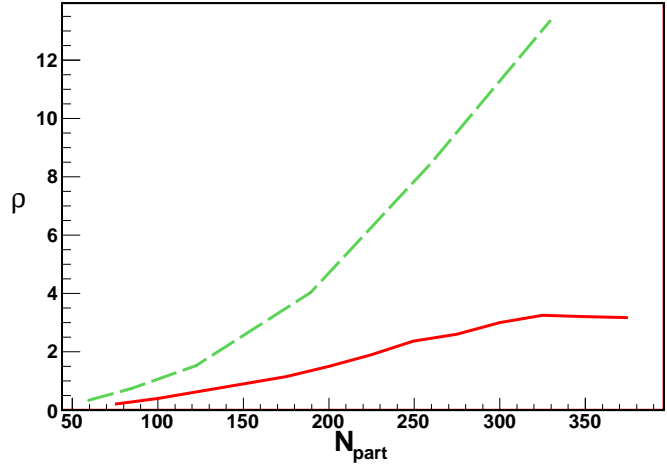}}
    \caption{String density dependence with the number of participants, 
dashed and solid lines correspond to $\sqrt{s}=2.76$ TeV and 
$\sqrt{s}=200$ GeV energies respectively.}
    \label{test4}
\end{center}
\end{figure}
\begin{figure}
\begin{center}
      \resizebox{100mm}{!}{\includegraphics{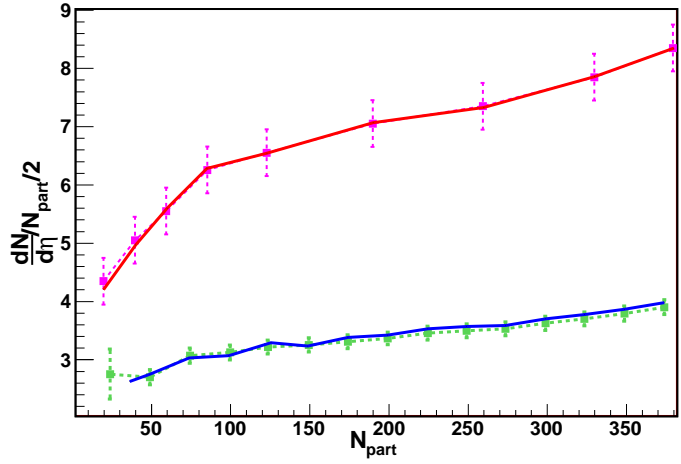}}
    \caption{lines in red and blue correspond to the dependence 
for the 
$\sqrt{s}=2.76$ TeV and $\sqrt{s}=200$ GeV energies, errorbars in green 
and pink correspond to ALICE data [22] and PHENIX data [31] respectively. }
    \label{test4}
\end{center}
\end{figure}

From the eq (13), we compute the transverse momentum dependence of
$v_{2}$ at different centralities. The results of $v_{2}$ at
different centralities are shown in figures (4) to (7), for
$Pb-Pb$ at $\sqrt{s}=2.76$ TeV compared to the RHIC results for
$Au-Au$ with ALICE and STAR data [30]. We obtain at all centralities
similar values of $v_{2}$ for $p_{T}<.8$ GeV/c and for $p_{T}$
larger the LHC values are slightly larger $(10\%)$ than RHIC
values. The difference is of the same size than the experimental
ones. Very similar results were obtained in the percolation
framework [20] using different approximations. In this case, the
values found are the same at RHIC and LHC in the whole $p_{T}$ 
range.
\begin{figure}
\begin{center}
      \resizebox{100mm}{!}{\includegraphics{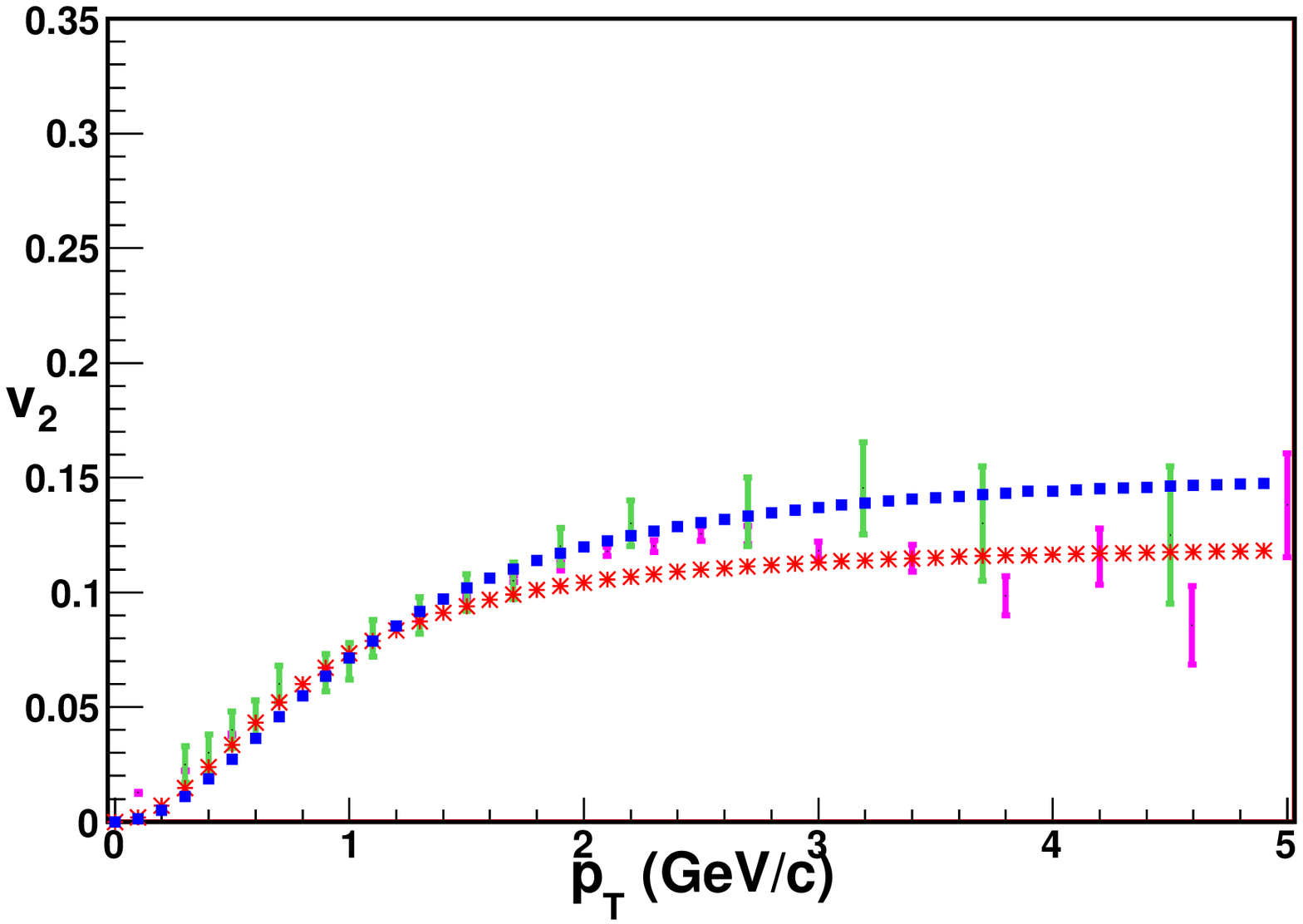}}
    \caption{Stars in red and blue squares correspond to our 
predictions for $\sqrt{s}=200$ GeV and $\sqrt{s}=2.76$ TeV
energies, and errorbars in green and pink are the respective 
data from references [22] [30] for centrality $10-20 \%$.}
    \label{test4}
\end{center}
\end{figure}
\begin{figure}
\begin{center}
      \resizebox{100mm}{!}{\includegraphics{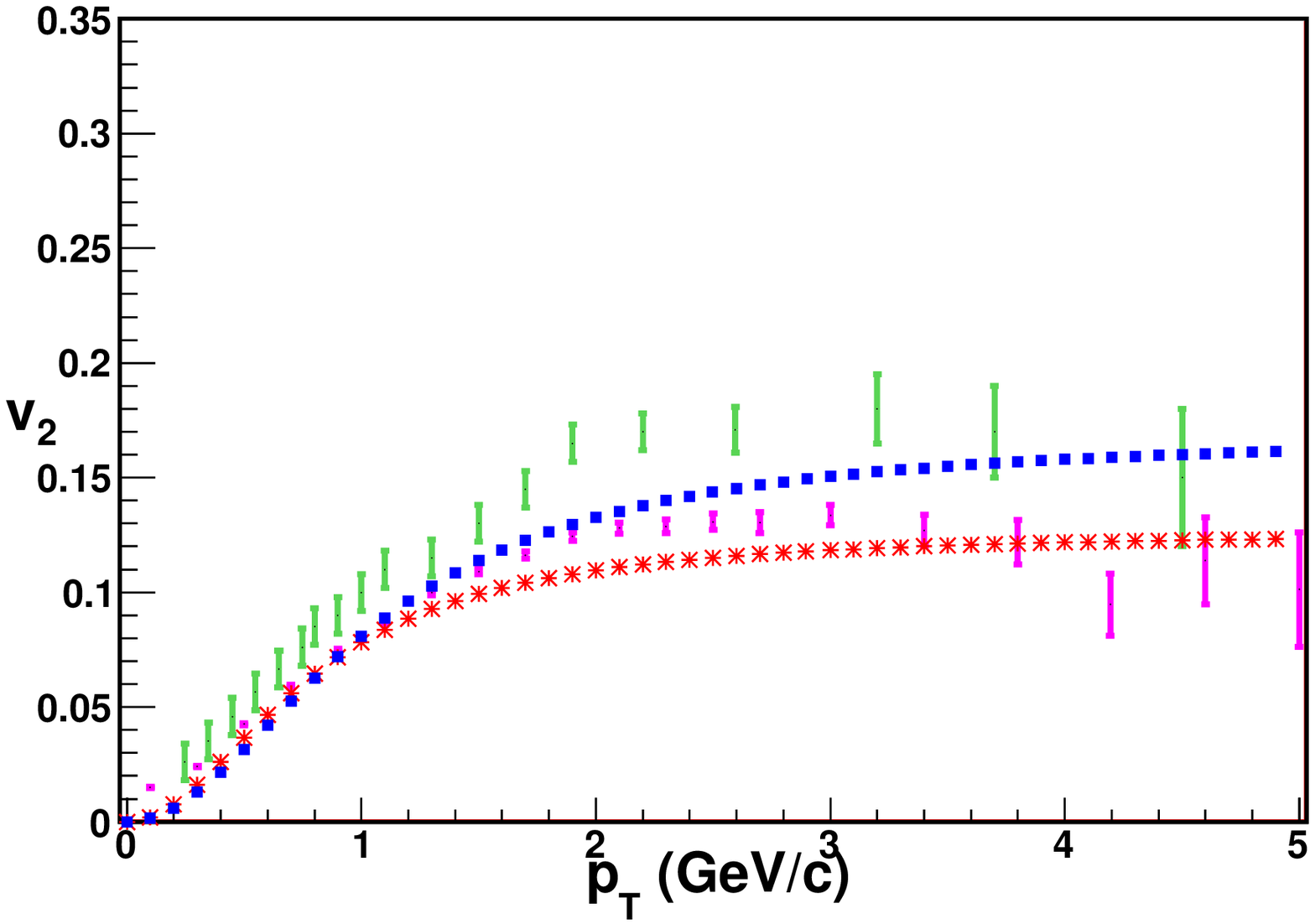}}
    \caption{Stars in red and blue squares correspond to our
predictions for $\sqrt{s}=200$ GeV and $\sqrt{s}=2.76$ TeV
energies, and errorbars in green and pink are the respective
data from references [22] [30] for centrality $20-30\%$.}
    \label{test4}
\end{center}
\end{figure}
\begin{figure}
\begin{center}
      \resizebox{100mm}{!}{\includegraphics{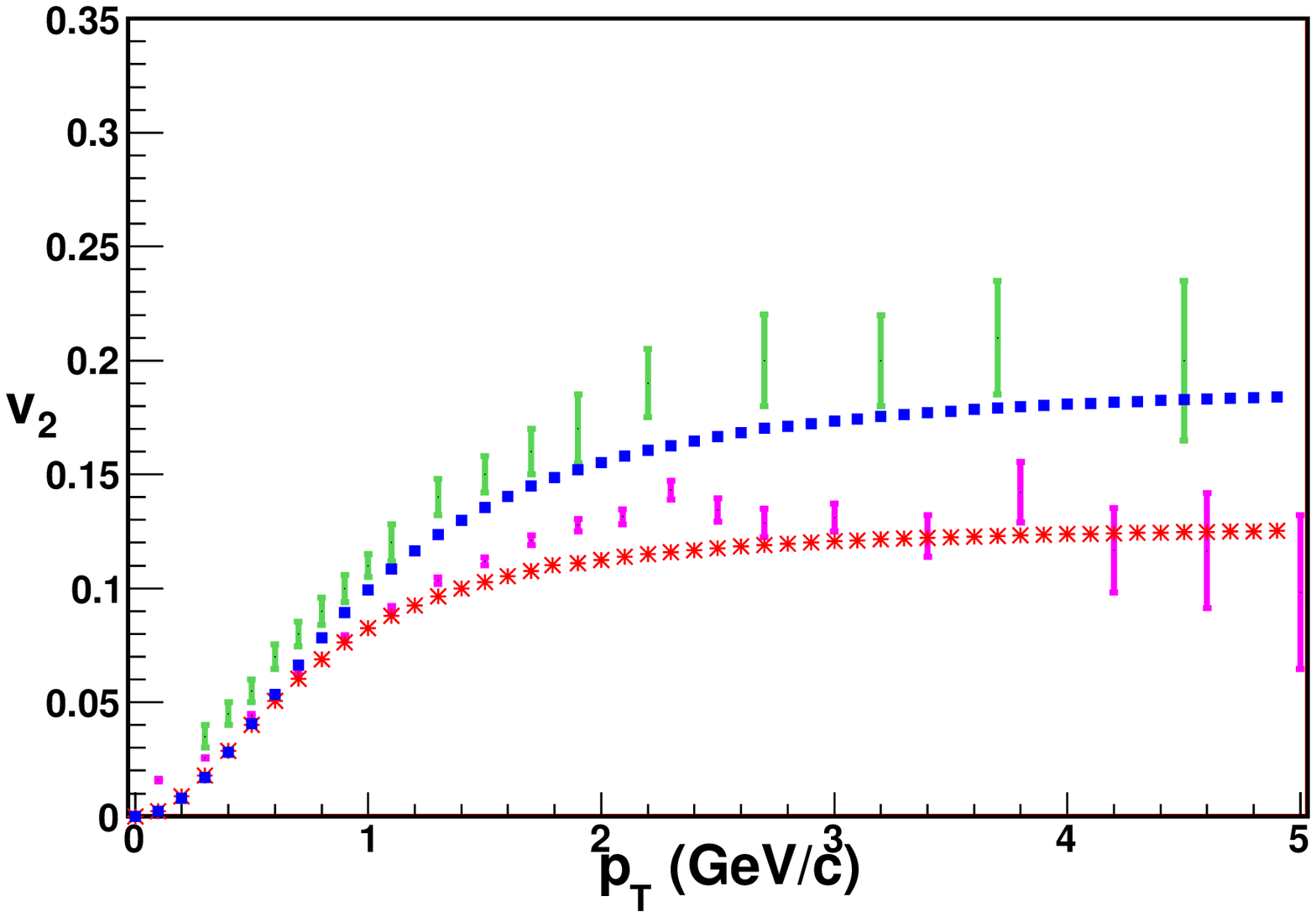}}
    \caption{Stars in red and blue squares correspond to our
predictions for $\sqrt{s}=200$ GeV and $\sqrt{s}=2.76$ TeV
energies, and errorbars in green and pink are the respective
data from references [22] [30] for centrality $30-40\%$.}
    \label{test4}
\end{center}
\end{figure}
\begin{figure}
\begin{center}
      \resizebox{100mm}{!}{\includegraphics{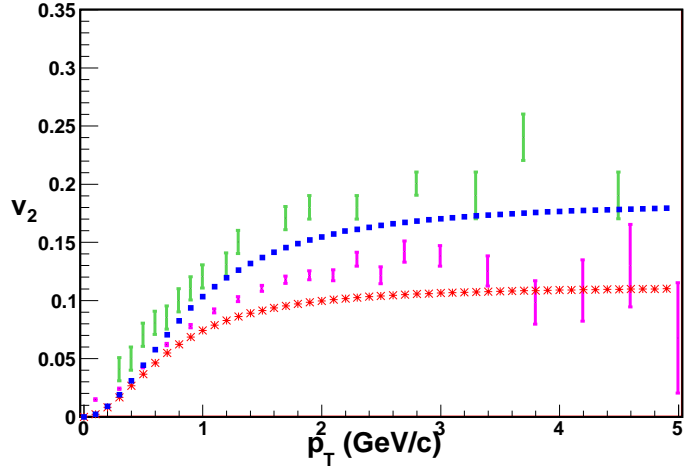}}
    \caption{Stars in red and blue squares correspond to our
predictions for $\sqrt{s}=200$ GeV and $\sqrt{s}=2.76$ TeV
energies, and errorbars in green and pink are the respective
data from references [22] [30] for centrality $40-50 \%$.}
    \label{test4}
\end{center}
\end{figure}

At $\sqrt{s}=5.5$ TeV, from the formula (13) we expect that
$v_{2}$ for low $p_{T}$ would be slightly smaller than at
$\sqrt{s}=2.76$ and almost equal at higher $p_{T}$.

In Fig. (8) and (9) we plot the rapidity dependence of $v_{2}$ 
for
central $Pb-Pb$ collisions at $\sqrt{s}=2.76$ TeV and for $Au-Au$
collisions at $\sqrt{s}=200$ GeV at $N_{pat}=211$ respectively,
obtained from eq(16). In Fig. (9), we include the corresponding
experimental data [32]. The integrated $v_{2}$ obtained is around
$25\%$ higher at LHC than at RHIC energies in agreement with the
experimental data [21]. We observe a breaking of limiting
fragmentation as it is expected for percolation in $\frac{dN}{dy}$
[33].
\begin{figure}
\begin{center}
      \resizebox{100mm}{!}{\includegraphics{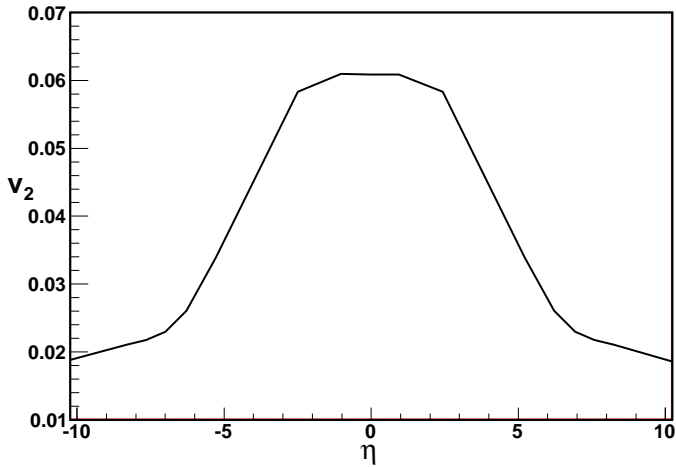}}
    \caption{Results for elliptic flow dependence with pseudorapidity in $Pb-Pb$ collisions at
 $\sqrt{s}=2.76$ TeV with  
$N_{part}=211$. }
    \label{test4}
\end{center}
\end{figure}
\begin{figure}
\begin{center}
      \resizebox{100mm}{!}{\includegraphics{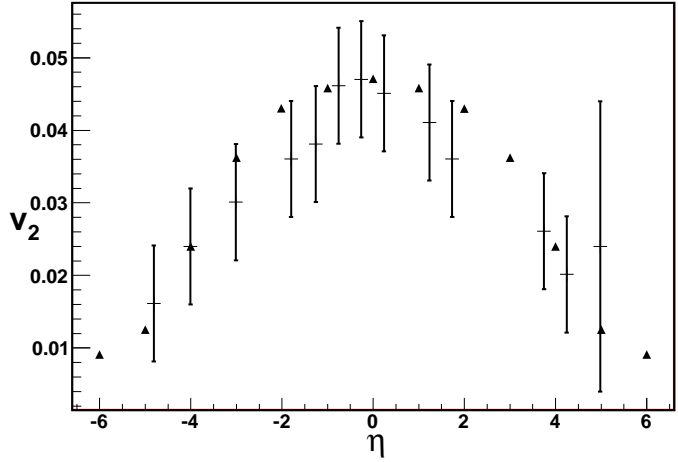}}
    \caption{Triangles show the corresponding results for elliptic flow 
dependence with pseudorapidity in   
$Au-Au$ 
collisions at $\sqrt{s}=200$ GeV with
$N_{part}=211$, compare to data from PHOBOS [32]. }
    \label{test4}
\end{center}
\end{figure}

In Fig. (10), we compare the values of $v_{2}$ at different 
centralities
with the experimental data, obtaining a good agreement.
 In fig (11) we show the results at $\sqrt{s}=2.76 $ TeV for 
$\pi$, $k$, 
and
$p$ at a centrality of $40-50\%$ for $Pb-Pb$ collisions. In
figure (12) we check the constituent quark scaling 
plotting the
ratio between the elliptic flow and the number of constituents as
a function of the ratio between the kinetic energy and the number
of constituents at $\sqrt{s}=2.76$ TeV. In this case, we 
observe an approximate scaling at low values of the kinetic energy. 
For high values, we have
deviations of the scaling at $\sqrt{s}=2.76$ TeV of the order of $20\%$. 
In our evaluations, we have neglected terms that for 
$p_{T}>1$ GeV/c can give contributions of the same order.
\begin{figure}
\begin{center}
      \resizebox{100mm}{!}{\includegraphics{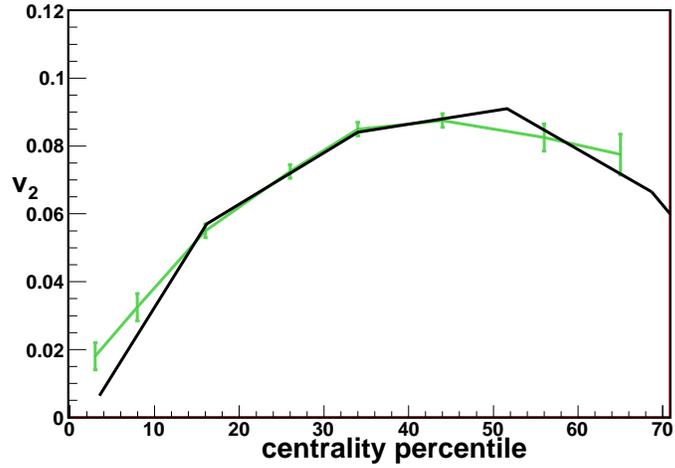}}
    \caption{Integrated $v_{2}$ at $\sqrt{s}=2.76$ TeV 
compared with ALICE data[22].}
    \label{test4}
\end{center}
\end{figure}
\begin{figure}
\begin{center}
      \resizebox{100mm}{!}{\includegraphics{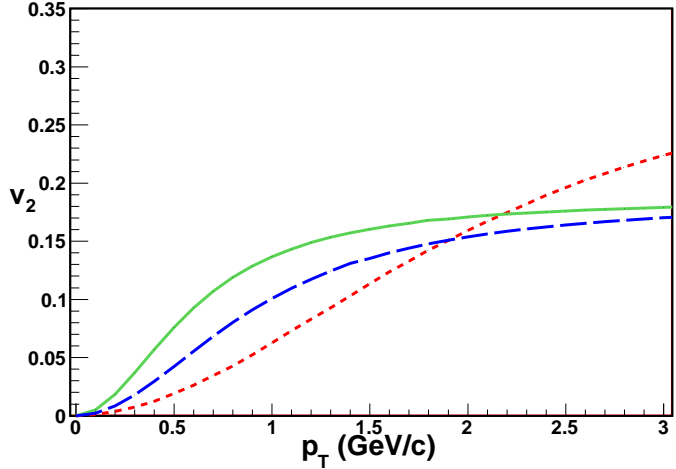}}
    \caption{Red doted line, green solid line and blue dashed line are 
correspond to the proton, kaon, and pion predictions for central $Pb-Pb$ 
collisions at $\sqrt{s}=2.76$ 
TeV.}
    \label{test4}
\end{center}
\end{figure}
\begin{figure}
\begin{center}
      \resizebox{60mm}{!}{\includegraphics{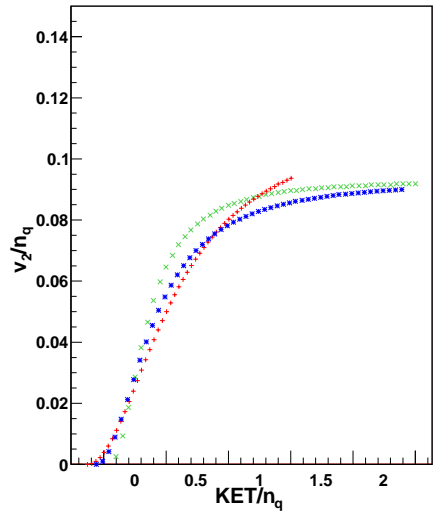}}
    \caption{Elliptic flow scaling with the number of quark 
contituents for central $Pb-Pb$ collisions at $\sqrt{s}=2.76$ TeV. 
Symbols ($+$) in red ,($*$) in blue and (x) in green  are used for protons 
kaons and pions respectively.}
    \label{test4}
\end{center}
\end{figure}

The similar values of $v_{2}$ found for LHC and RHIC energies
points out the existence of a liquid of very low value of the
ratio between the shear viscosity $\eta$ and the entropy, $s$. This low 
ratio
is expected in percolation approach [23]. In fact, the ratio is
given by
\begin{equation}
\frac{\eta}{s}\simeq\frac{T}{5 n \sigma_{tr}}
\end{equation}
where $T$ is the temperature, $n$ the number of effective sources per
unit volume and $\sigma_{tr}$ is the transport cross section,
According to the second part of the formula (5), the transverse
correlation length is $r_{0}\sqrt{F(\rho)}$ and therefore
\begin{equation}
\sigma_{tr}=\pi r_{0}^{2} F(\rho)=S_{1}F(\rho)
\end{equation}
On the other hand, the number of sources is given by the ratio
between the area covered by strings and the area corresponding to
the transverse correlation length
\begin{equation}
N=\frac{(1-e^{-\rho})S_{A}}{F(\rho)S_{1}}
\end {equation}
and therefore
\begin{equation}
n=\frac{(1-e^{-\rho})}{F(\rho)S_{1}L}
\end{equation}
with $L\simeq 1$ fm.

Finally, we can relate the temperature to the tension of the
cluster [16] resulting
\begin{equation}
T=(<p_{T}^{2}>_{1}/2F(\rho))^{1/2}
\end{equation}

Introducing (19), (20), (21) and (22) into (18) we obtain
\begin{equation}
\frac{\eta}{s}=\frac{<p_{T}>_{1}L}{5\sqrt{2}}
\frac{\rho^{1/4}}{(1-e^{-\rho})^{5/4}}
\end{equation}
where we have approximated $(<p_{T}^{2}>_{1})^{1/2}\simeq
<p_{T}>_{1}$. For $<p_{T}>_{1}=200$ MeV/c,
 close to the threshold of
percolation $\rho\simeq 1.2-1.5$, we obtain $\eta/s \simeq 0.2$
and for $\rho\simeq 3-5$ we obtain $\eta /s\simeq .25$. The
ratio grows slowly, as $\rho^{1/4}$ remaining low between RHIC and
LHC energies. Let us remark, that this low ratio is due to the
small transverse correlation length, decreasing as the energy
increases compensating the increasing of the effective number of
sources in such a way that the mean path length
$\lambda=1/n \sigma=\frac{L}{ (1-e^{-\rho}) }$ remains almost
constant.

\section{Conclusions}
 The close analytical formulas obtained in the
percolation framework from the approximated transverse momentum
distribution allow us to compute the elliptic flow once the string
density is known, which is fixed from the experimental data at LHC
and RHIC energy. The transverse momentum dependence of $v_{2}$ and
its values are very similar at RHIC and LHC energies for all
centralities, in agreement with the experimental data. The
integrated elliptic flow at mid rapidity is around $25\%$ higher at
$\sqrt{s}=2.76$ TeV than at $\sqrt{s}=200$ GeV and in the rapidity
dependence of the elliptic flow there is no longer a triangle
shape shown at RHIC. In the percolation framework there is not
limiting fragmentation for both $v_{2}$ and $dN/dy$. The whole
picture is consistent with the formation of a fluid with a low ratio 
between shear viscosity and entropy in the range of energies of RHIC and
LHC such as it is expected in percolation. The percolation framework 
provide us with a microscopic partonic picture which explains the early 
thermalization and the large interaction required by the hydrodynamical 
approach.
\\
\\
\\
\begin{Large}{Acknowledgments.}\end{Large}\\

We thank B. K. Srivastava, M. A. Braun, N. Armesto and C. Salgado, for 
useful discussions.

J. D. D. thanks the support of the FCT/Portugal project 
PPCDT/FIS/575682004.

I. B, and C. P. were supported by the project FPA2008-01177 of
MICINN, the Spanish Consolider –Ingenio 2010 program CPAN and Conselleria Educacion Xunta de Galicia.

 \newpage
\begin{Large}{References}\end{Large}

\end{document}